\documentclass[11pt,a4paper]{article}
\usepackage[utf8]{inputenc}
\usepackage[T1]{fontenc}
\usepackage{amsmath,amssymb,amsthm}
\usepackage{mathptmx}
\usepackage[scaled=0.92]{helvet}
\usepackage{courier}
\usepackage{booktabs}
\usepackage{array}
\usepackage{graphicx}
\usepackage{placeins}
\usepackage[expansion=false]{microtype}
\usepackage[top=25mm,bottom=25mm,left=27mm,right=27mm]{geometry}
\usepackage[font=small,labelfont=bf]{caption}
\PassOptionsToPackage{hyphens}{url}
\usepackage[colorlinks=true,linkcolor=blue,citecolor=blue,urlcolor=blue]{hyperref}
\theoremstyle{remark}
\newtheorem{remark}{Remark}

\newcommand{\nEB}{n_{\mathrm{EB}}}
\newcommand{\lmin}{\lambda_{\min}}
\newcommand{\kmin}{k_{\min}}
\newcommand{\Tr}{\mathrm{Tr}}
\newcommand{\doilink}[1]{\href{https://doi.org/#1}{doi:#1}}
\newcommand{\arxivlink}[1]{\href{https://arxiv.org/abs/#1}{arXiv:#1}}

\begin{document}

\begin{center}
{\LARGE\bfseries Pulsed control of an entanglement-breaking index:\\[3pt]
thermal valleys, a certified valve,\\[3pt]
and its open-close asymmetry\par}
\vspace{14pt}
{\large Eran Kopel\par}
\vspace{4pt}
Tel Aviv University, Tel Aviv, Israel\\[2pt]
\href{mailto:erankopel@tauex.tau.ac.il}{erankopel@tauex.tau.ac.il}\quad
ORCID \href{https://orcid.org/0000-0003-4657-8636}{0000-0003-4657-8636}\\[8pt]
\today
\end{center}

\vspace{4pt}
\begin{center}
\begin{minipage}{0.92\textwidth}
\small
\noindent\textbf{Abstract.}
The entanglement-breaking index of a channel is the number of self-compositions after
which all entanglement with a reference system is destroyed. In a thermal qubit collision
loop whose round is affine in the bath polarisation $p$, some circuits
have thermal valleys: on an interior window of $p$ the index falls below its
infinite-temperature value. Rounds at different $p$ compose exactly, so every schedule of
polarisations defines a composite channel whose entanglement-breaking status can be
certified in ball arithmetic. A gated search of $33$ million random circuits finds no valley with contrast
above about $3\times10^{-3}$, with an interior maximum near index $90$, and yields six
certified targets. At the reference target, held inside its window for $88$ rounds, a
centred excursion to $p = 0.05$ restores entanglement at exactly eight rounds: every
shorter centred pulse certifiably fails, and so do the same eight rounds at the start.
A $53$-round pulse certifiably beats the static protocol, a
depth-three valley is walked down its staircase by pulse duration, and a $1.2$ per cent
coherent tilt of the ancilla opens the switch at fixed population. Closing the switch
costs $21$ to $35$ rounds. The asymmetry is not one of local rates, which agree to within
about two per cent at matched circuit positions (certified). It is one of distance and
response: closing must remove an entanglement margin $2.2$ times larger, and while hot
rounds added to a window hold act almost independently, window rounds added to a hot hold
act collectively. Near the crossing, the lever of the log-margin is that of a linear
approach to the entanglement-breaking boundary, with no free parameter. At the reference
target, depolarising noise of about $4\nu/n$ per round kills a signature of size $\nu$
read after $n$ rounds: $1.3\times10^{-4}$ for the static signal, $6.5\times10^{-6}$ for
the eight-round switch.
\end{minipage}
\end{center}

\vspace{6pt}

\section{Introduction}\label{sec:intro}

A channel is entanglement breaking if its output is separable from any reference system
for every input \cite{HSR2003}. A channel $\Phi$ that is not entanglement breaking can
become so under composition, and the least such power,
\begin{equation}
  \nEB(\Phi) \;=\; \min\{\, n \ge 1 : \Phi^n \text{ is entanglement breaking}\,\},
  \label{eq:index}
\end{equation}
is its entanglement-breaking index \cite{DePasquale2012,LG2015}. Maps that become
entanglement breaking after finitely many uses, and the number of uses they need, have
been characterised in increasing generality
\cite{Rahaman2018,Christandl2019,Hanson2020,Park2026}. For a qubit channel the index is
decided by a finite computation: $\Phi^n$ is entanglement breaking exactly when the
partial transpose of its Choi matrix \cite{Choi1975} is positive semidefinite
\cite{Peres1996,Horodecki1996,HSR2003}.

Collision models describe open dynamics as repeated interactions of a system with fresh
ancillas \cite{Scarani2002,Ciccarello2022,Campbell2021}, so one collision is one channel
and the index counts collisions. The loop studied here is one round of the three-qubit
feedback circuit of Ref.~\cite{PaperI}, in which a message qubit meets two fresh
ancillas, here prepared in a thermal state as in Ref.~\cite{PaperII}. That companion work \cite{PaperII} decomposed its index into an
architecture floor, the index at infinite temperature, and a thermal term, and exhibited
certified circuits whose index dips below the floor on an interior window of the bath
polarisation: a bath of intermediate temperature breaks entanglement sooner than either
extreme. We call such circuits thermal valleys. This paper asks what a valley can do
when the polarisation is not held fixed but changed from round to round.

The question is tractable because each round acts affinely on the Bloch vector of the
message qubit, with coefficients that are polynomials in the polarisation. A schedule of
rounds at different polarisations therefore composes exactly into another affine pair, and
its Choi matrix and partial-transpose spectrum can be enclosed rigorously. A schedule is
as certifiable as a static power.

The results are of four kinds.
\begin{enumerate}
\item \emph{The wall.} A gated random search of $33$ million circuits finds no valley
      whose contrast, the smallest of the signals that separate the dip from its
      surroundings, exceeds a few times $10^{-3}$. Contrast rises with the floor, peaks
      near floor $90$ at a certified $2.85\times10^{-3}$, and declines. Six certified
      targets carry the rest of the paper (Sec.~\ref{sec:valley}).
\item \emph{The valve.} Held inside its window, the reference target breaks all
      entanglement by its read-out round. A centred excursion out of the window for
      exactly eight rounds restores it, with every shorter centred pulse certified not
      to and the same pulse at the start certified not to; a longer, $53$-round pulse
      gives a certified $20$ per cent more negativity than the static protocol; and a
      certified depth-three valley is walked down its staircase by pulse duration
      (Sec.~\ref{sec:valve}).
\item \emph{The asymmetry.} Closing the valve costs $21$ to $35$ rounds where opening
      costs $8$. We show that this is not a difference of local rates, which agree at
      matched circuit positions to within about two per cent, but of distance and
      response, and that near the crossing the lever of the log-margin follows, with no
      fitted parameter, from a linear approach to the entanglement-breaking boundary
      (Sec.~\ref{sec:asym}).
\item \emph{Coherence and noise.} A tilt of the ancilla Bloch vector by $1.2$ per cent,
      with the thermal population held fixed, opens the valve; sustained tilts raise the
      static signal about $2.5$-fold and a centred $26$-round tilt triples it
      (Sec.~\ref{sec:tilt}). Within an exact depolarising model a signature of the
      reference target of size $\nu$ dies at per-round noise $\eta^\ast \approx 4\nu/n$:
      near $10^{-4}$ for the static signal and the strongest pulses, the best of which
      outlasts the static protocol, but at $6.5\times10^{-6}$ for the eight-round switch
      (Sec.~\ref{sec:noise}).
\end{enumerate}

Every number called \emph{certified} is a two-sided enclosure or a proven integer in
ball arithmetic at $256$ bits \cite{Arb2017,pythonflint}, obtained from one-sided
primitives only: a negative eigenvalue is proved by an explicit witness vector, positive
semidefiniteness by Sylvester's criterion. The margins certified here, from about
$10^{-6}$ at the window edges to about $10^{-2}$, are far above double-precision rounding, so floating-point arithmetic gives
the same signs; ball arithmetic turns those signs into proofs at negligible cost, which
matters for statements at a boundary such as \emph{exactly eight rounds}. Numbers from
floating-point searches and scans are labelled \emph{float64} and are never promoted.

All results concern a model channel. Measuring a valley signature on hardware requires
of order $10^{8}$ shots, circuits of $80$ to $110$ rounds with mid-circuit reset, and an
effective per-round error of $10^{-4}$ or less; a measurement protocol for the index of this
loop, aimed at targets that are measurable today, is the subject of version~3 of
Ref.~\cite{PaperV}. Versions~1 and~2 of that record contained an earlier form of the
valley, pulse and tilt analysis; version~3 keeps only a short summary of it. The present
paper supersedes those sections, completes the certificate of the minimal pulse, and adds the analysis of the
asymmetry. Section~\ref{sec:claims} lists what is claimed and at what level.

\section{The channel, schedules, and certification}\label{sec:model}

\subsection{One round}\label{sec:round}

One round acts on the message qubit $M$ and two fresh ancillas $F$ and $L$, each
prepared in the thermal state
\begin{equation}
  \tau(p) \;=\; \tfrac12\bigl(\mathbb{I} + p\,Z\bigr), \qquad p \in [0, 1],
  \label{eq:tau}
\end{equation}
where $p = \tanh(\hbar\omega/2kT)$ is the polarisation of a two-level bath of splitting
$\hbar\omega$ at temperature $T$: $p = 0$ is infinite temperature and $p = 1$ is zero
temperature. With registers ordered $(M, F, L)$ and $R_y(a) = \exp(-iaY/2)$, the round
unitary is
\begin{equation}
  U(\theta, \varphi, \kappa, \beta) \;=\; \bigl(R_y(\beta)\otimes\mathbb{I}\otimes\mathbb{I}\bigr)\,
  U_f(\varphi)\, U_w(\kappa)\, U_W(\theta),
  \label{eq:U}
\end{equation}
with
\begin{align}
  U_W(\theta) &= |0\rangle\langle 0|_M \otimes R_y(\pi - 2\theta)_F \otimes \mathbb{I}_L
               + |1\rangle\langle 1|_M \otimes R_y(2\theta)_F \otimes \mathbb{I}_L,\\
  U_w(\kappa) &= \exp\bigl(-i\tfrac{\kappa}{2}\, Z_F \otimes Y_L\bigr),\qquad
  U_f(\varphi) = \exp\bigl(-i\varphi\,\mathrm{SWAP}_{ML}\bigr).
\end{align}
$U_W$ writes the message into the feedback ancilla $F$, $U_w$ couples $F$ to the leak
ancilla $L$, and the partial swap $U_f$ returns part of the leaked information to $M$.
The reading of this circuit as a feedback loop is the subject of Ref.~\cite{PaperI};
here it is a family of qubit channels,
\begin{equation}
  \Phi_p(\rho) \;=\; \Tr_{FL}\bigl[\,U\bigl(\rho\otimes\tau(p)\otimes\tau(p)\bigr)U^\dagger\bigr].
  \label{eq:channel}
\end{equation}
All targets of Table~\ref{tab:targets} have leak strength $\varepsilon = \sin^2\varphi =
0.002$, that is $\varphi = 0.044736280102247346$, and differ in $(\theta, \kappa, \beta)$.

\subsection{Affine representation and schedules}\label{sec:affine}

Writing $\rho = \tfrac12(\mathbb{I} + \mathbf r\cdot\boldsymbol\sigma)$, a qubit channel acts
as $\mathbf r \mapsto A\mathbf r + \mathbf c$ with a real $3\times3$ matrix $A$ and a real
vector $\mathbf c$. For the round \eqref{eq:channel} each ancilla enters linearly through
$p$, so the entries of $A(p)$ and $\mathbf c(p)$ are quadratic polynomials in $p$. A
\emph{schedule} $\mathbf p = (p_1, \dots, p_n)$, applied in that order, composes exactly:
\begin{equation}
  A_{\mathbf p} = A(p_n)\cdots A(p_1),\qquad
  \mathbf c_{\mathbf p} = \sum_{j=1}^{n} A(p_n)\cdots A(p_{j+1})\, \mathbf c(p_j),
  \label{eq:compose}
\end{equation}
and a constant schedule gives the power $\Phi_p^n$. Per-round depolarising noise of
strength $\eta$ on $M$ acts exactly as $(A, \mathbf c) \to (1-\eta)(A, \mathbf c)$
per round, so noisy schedules are again affine pairs.

Entanglement breaking is decided on the Choi matrix
$J(\Phi) = \tfrac12\sum_{i,j}|i\rangle\langle j|\otimes\Phi(|i\rangle\langle j|)$
\cite{Choi1975}: $\Phi$ is entanglement breaking exactly when $J(\Phi)$ is separable
\cite{HSR2003}, which for two qubits is positivity of the partial transpose
\cite{Peres1996,Horodecki1996}. Let $H$ be the partial transpose of $J$ on the reference
factor and $\lmin$ its least eigenvalue. A two-qubit partial transpose has at most one
negative eigenvalue \cite{Sanpera1998}, so the negativity \cite{VidalWerner2002} of the
Choi state is $\max(0, -\lmin)$, and we use $\lmin$ itself, of either sign, as the
signal: negative means entangled (NPT), nonnegative means entanglement breaking (PPT).
Since entanglement-breaking channels form an ideal under composition, a schedule whose
first $r$ rounds compose to a PPT channel stays PPT for every later round. Along any
schedule, the sign of $\lmin$ of the first $r$ rounds can therefore change only once,
from negative to nonnegative; we call the round at which it changes the crossing.

\subsection{Index, floor, valley, contrast}\label{sec:defs}

For a fixed circuit, write $H_n(p)$ for the partial transpose of the Choi matrix of
$\Phi_p^n$ and use the following terms.
\begin{itemize}
\item The \emph{index} $\nEB(p) = \min\{n : \lmin[H_n(p)] \ge 0\}$, and the \emph{floor}
      $F = \nEB(0)$, the index at infinite temperature.
\item The \emph{grid}: sixteen polarisations $p_j = 0.05 + 0.95\,j/15$, $j = 0, \dots, 15$.
\item A \emph{valley}: a circuit whose index at some grid point is below its floor. Its
      \emph{window} $W$ is the set of grid points with $\nEB(p) < F$. The valley is a
      \emph{true return} if $W$ excludes $p = 1$ and a \emph{cliff} if it does not. Its
      \emph{depth} is $F - \min_p \nEB(p)$.
\item The \emph{contrast} of a valley,
      \begin{equation}
        C = \min(m_{\mathrm{in}}, \nu_{0.05}, \nu_1),\quad
        m_{\mathrm{in}} = \max_{p\in W}\lmin[H_{F-1}(p)],\quad
        \nu_{q} = -\lmin[H_{F-1}(q)],
        \label{eq:contrast}
      \end{equation}
      with $\nu_1$ omitted for a cliff. At $n = F - 1$ rounds the channel is entangled at
      both ends of the grid and entanglement breaking inside $W$, and $C$ is the smallest
      of the three signals that distinguish the dip from its surroundings.
\end{itemize}

\subsection{Certification}\label{sec:cert}

Circuit parameters are the binary64 numbers drawn by the search, printed in
Table~\ref{tab:targets} as their shortest decimal representations. Grid polarisations are the
binary64 values of NumPy's \texttt{linspace(0.05, 1, 16)}, within one unit in the last
place of $0.05 + 0.95\,j/15$, printed rounded (for example $0.367$, or $p^\ast = 0.5567$
for D3); every other polarisation is the decimal printed. Depending on
the script, a certificate takes a binary64 number or its shortest decimal, which differ by
less than $10^{-16}$, and it concerns the circuit so specified; no certified margin comes
within many orders of magnitude of that difference. Claims are established in Arb ball
arithmetic \cite{Arb2017} through python-flint \cite{pythonflint} at $256$ bits, by
one-sided primitives only. A negative eigenvalue of $H$ is proved by a witness vector $v$
whose Rayleigh quotient $v^\dagger H v/v^\dagger v$ is certified negative; this quotient
is a rigorous upper bound on $\lmin$ for any $v$. Positive semidefiniteness is proved by
Sylvester's criterion on the leading principal minors. A two-sided enclosure of $\lmin$
combines the two: the Rayleigh quotient of a floating-point eigenvector gives the upper
bound, and Sylvester's criterion on $H - s\,\mathbb{I}$ certifies $\lmin > s$, with $s$
backed off from the floating-point value until the shifted certificate passes. Enclosures
quoted below have relative width below $10^{-9}$, and we print at most eight significant
digits. Logarithmic quantities, such as the lever of
Sec.~\ref{sec:asym}, are evaluated on these enclosures by ball arithmetic.

Three labels are used. \emph{Certified}: a two-sided enclosure or a proven sign or
integer as above. \emph{Float64}: output of searches and scans, reported as statistics.
\emph{Arithmetic}: shot counts and similar figures derived from stated formulas. The
certification pipeline of the searches first reproduces a reference affine pair to twenty
significant digits, and the certification scripts include planted false claims (an NPT
claim at a certified PPT point, a PPT claim at a certified NPT point, a shift above a
certified Rayleigh bound) that the primitives are shown to reject. The schedule engine is
checked before use: a constant schedule reproduces the matrix-power route, and an early
and a late half-pulse differ by up to $0.19$ in individual entries of $A_{\mathbf p}$, so
the engine sees the non-commutativity that the control results exploit. Appendix~\ref{app:files} lists the
scripts that re-derive every certified number from these primitives.

\section{Thermal valleys: the wall and the certified targets}\label{sec:valley}

\subsection{The search}\label{sec:search}

A valley hovers at the entanglement-breaking boundary, so its signals are small by
nature. The search was designed around one question, fixed before it ran: do valleys
with contrast $C \ge 0.05$ exist, $0.05$ being roughly what two-qubit tomography
resolves at $10^4$ shots per setting against realistic noise? Either answer was to be
reported. The answer is no, and the shape of the no is the finding.

Circuits were drawn at random with the leak strength in bins
$\varepsilon = \sin^2\varphi$ from $0.002$ to $0.125$, $\theta$ normal about $\pi/4$
with standard deviation $0.05\pi$, and $\kappa$, $\beta$ uniform on $[0, \pi/2]$. Three
stratified searches ran on a sharded, journalled chassis, each with a closed-form
prescreen of the floor, the full sixteen-point index profile of every kept circuit, and
the contrast of every valley: floors $2$ to $30$ ($24$ million draws, eight bins of
$\varepsilon$), floors $31$ to $60$ ($6$ million draws, $\varepsilon = 0.002$ and
$0.005$), and floors $61$ to $120$ ($3$ million draws, $\varepsilon = 0.002$), in all
$33$ million draws and $10{,}628$ valleys. Each run was gated on the unitarity of the
embedded circuits, on agreement of the quadratic-in-$p$ reconstruction with direct
evaluation at every grid point, on agreement of the closed-form floor with the
Choi-matrix floor, and on end-to-end reproduction of a reference valley including its
contrast. Each regime was entered
through a certified anchor at its own boundary: a floor-$3$ reference valley with window
$[0.177, 0.493]$ and certified contrast $2.9328\times10^{-5}$, then a certified valley
from each search as the reference of the next, $5.6400771\times10^{-4}$ (a cliff of floor
$30$) and $1.1509431\times10^{-3}$ (a true return of floor $60$). The search output is
float64; everything quoted as certified below was recomputed as an enclosure.

\subsection{The shape of the wall}\label{sec:wall}

No valley found in $33$ million draws reaches the threshold. The strongest object of any
kind is a cliff with certified $C = 3.0646104\times10^{-3}$, a factor $16$ below it; the
strongest true return has certified $C = 2.8502660\times10^{-3}$. This is an empirical
statement about a gated search of stated size with certified extreme points, not a
non-existence theorem. What the search does establish is a shape (Table~\ref{tab:wall}):
the contrast of true returns rises with the floor across seven bands, peaks at floors
$86$ to $95$, and declines, with comparable samples in every band, so the bend is
structure rather than sampling. A power law fitted to the rising flank of the first two
searches would have placed $C = 0.05$ far beyond the probed range, at floors between about
$340$ and $1{,}300$ depending on the statistic fitted (float64); the third search
shows that the rise bends over inside the probed range. Two further facts are
structural: no cliff occurs among the $1{,}354$ valleys with floors $106$ to $120$, and
the $90$th percentile of the contrast per collision, $C/(F-1)$, peaks in the same band
as that of the contrast itself.

\begin{table}[t]
\centering
\begin{tabular}{lcc@{\qquad}lcc}
\toprule
floors & $C$ (q90) & $C$ (max) & floors & $C$ (q90) & $C$ (max)\\
\midrule
$26$--$35$  & $1.9\times10^{-4}$ & $3.9\times10^{-4}$ & $76$--$85$   & $1.56\times10^{-3}$ & $2.74\times10^{-3}$\\
$36$--$45$  & $3.4\times10^{-4}$ & $6.5\times10^{-4}$ & $86$--$95$   & $\mathbf{1.81\times10^{-3}}$ & $\mathbf{2.85\times10^{-3}}$\\
$46$--$55$  & $4.7\times10^{-4}$ & $9.7\times10^{-4}$ & $96$--$105$  & $1.76\times10^{-3}$ & $2.66\times10^{-3}$\\
$56$--$65$  & $6.5\times10^{-4}$ & $1.58\times10^{-3}$ & $106$--$115$ & $1.63\times10^{-3}$ & $2.44\times10^{-3}$\\
$66$--$75$  & $1.06\times10^{-3}$ & $2.20\times10^{-3}$ & $116$--$125$ & $1.59\times10^{-3}$ & $2.17\times10^{-3}$\\
\bottomrule
\end{tabular}
\caption{Contrast of true-return valleys by floor band, pooled over the three searches
(float64; per-band $90$th percentile and maximum; $235$ to $1{,}046$ valleys per band).
The maximum is interior. Certified extreme points: $2.8502660\times10^{-3}$ at floor $89$
(true return) and $3.0646104\times10^{-3}$ at floor $74$ (cliff).}
\label{tab:wall}
\end{table}

We record the observation as a conjecture: over this architecture, valley contrast is
bounded by a few times $10^{-3}$. A valley is a boundary phenomenon, and the conjecture
says that its visibility cannot exceed its distance from the entanglement-breaking set;
a proof or a counterexample would be a statement about the geometry of that set for the
collision family.

\subsection{Certified targets}\label{sec:targets}

Six certified circuits carry the rest of the paper (Table~\ref{tab:targets}). For each
of the five contrast targets, the floor is fixed by a witness certificate at $n = F - 1$
and a Sylvester certificate at $n = F$, both at $p = 0$; the sign of $\lmin[H_{F-1}(p)]$
is certified at all sixteen grid points, consistent with the window (an enclosure that
straddles zero counts as a failed certificate, never as a pass); and $m_{\mathrm{in}}$,
$\nu_{0.05}$, $\nu_1$ and $C$ are enclosed from both sides. For D3 the floor and one
interior index are certified.

\begin{table}[t]
\centering
\small
\setlength{\tabcolsep}{5pt}
\begin{tabular}{lrllll}
\toprule
target & floor & window (grid) & type & certified $C$ & role\\
\midrule
QR1 & $89$  & $[0.367, 0.873]$ & true return & $2.8502660\times10^{-3}$ & reference target\\
TR2 & $86$  & $[0.367, 0.873]$ & true return & $2.8091775\times10^{-3}$ & runner-up\\
TR3 & $89$  & $[0.367, 0.873]$ & true return & $2.7748002\times10^{-3}$ & runner-up\\
QC1 & $81$  & $[0.367, 0.873]$ & true return & $2.7408919\times10^{-3}$ & best true-return $C/(F-1)$\\
QCL & $74$  & $[0.430, 1]$     & cliff       & $3.0646104\times10^{-3}$ & strongest cliff\\
D3  & $109$ & $[0.240, 0.810]$ & true return & depth $\ge 3$ (integers) & depth example\\
\bottomrule
\end{tabular}

\vspace{6pt}
\footnotesize
\setlength{\tabcolsep}{6pt}
\begin{tabular}{llll}
\toprule
target & $\theta$ & $\kappa$ & $\beta$\\
\midrule
QR1 & $0.8762894250210506$ & $0.1247057659254267$  & $0.05210240689909829$\\
TR2 & $0.6927332755353641$ & $0.12421594217334223$ & $0.05927844940712698$\\
TR3 & $0.8762903871107547$ & $0.13875023182455581$ & $0.049803656670883066$\\
QC1 & $0.6891907806455668$ & $0.11191302043215262$ & $0.05538719740686704$\\
QCL & $0.8867922385265417$ & $0.2614659298771744$  & $0.06582547774707999$\\
D3  & $0.8658200621431981$ & $0.2603365388964298$  & $0.04415909934609924$\\
\bottomrule
\end{tabular}
\caption{Certified valley targets (top) and their parameters (bottom; the binary64
values drawn, as shortest decimals; $\varphi = 0.044736280102247346$ for all). Floors and window sign patterns are
certified integers and signs, contrasts certified two-sided enclosures. For QR1,
$m_{\mathrm{in}} = 3.626\times10^{-3}$ and $\nu_1 = 2.943\times10^{-3}$, so the binding
term is $\nu_{0.05} = 2.8502660\times10^{-3}$. For D3 the certified statements are the
floor $109$ and the index exactly $106$ at $p^\ast = 0.5567$; its window is float64.}
\label{tab:targets}
\end{table}

\textbf{QR1}, the reference target, has floor $89$ and certified contrast
$2.8502660\times10^{-3}$, $97$ times that of the floor-$3$ reference valley; its
runners-up TR2 and TR3 show a family rather than an outlier. \textbf{QC1} has the best
contrast per collision among the certified true returns, $C/(F-1) = 3.43\times10^{-5}$,
with eight fewer rounds than QR1 for a four per cent loss of contrast. \textbf{QCL}, the
strongest cliff, lacks a return; it has the largest contrast found. \textbf{D3} is a
certified depth-three valley: its floor is exactly $109$ and its index at
$p^\ast = 0.5567$ exactly $106$, so its depth is at least $3$ (exactly $3$ in float64).
Deeper certified valleys exist at larger floors: the companion work certifies a circuit
with floor $352$ and index $347$ at $p = 0.5$ \cite{PaperII}. Depth grows with the floor:
none of the $214$ valleys with floors up to $30$ is deeper than $1$; four of the
$3{,}820$ with floors $31$ to $60$ have depth $2$ (one certified to depth at least $2$);
and $1{,}589$ of the $6{,}594$ with
floors $61$ to $120$, $24$ per cent, have depth at least $2$, $142$ of them depth $3$
(float64 statistics).

Bisection with certified brackets on each side (a PPT certificate inside, an NPT witness
outside) fixes the true window of QR1 beyond grid resolution:
\begin{equation}
  \text{left edge} \in (0.3175,\ 0.3177), \qquad \text{right edge} \in (0.9137,\ 0.9138).
  \label{eq:edges}
\end{equation}
The hold polarisation $p_{\mathrm h} = 0.4016$ used below is an interior point of this
window, $0.084$ above its left edge; it has no other significance, and every certificate
of Secs.~\ref{sec:valve} to \ref{sec:noise} is issued at it.

\section{Pulsed schedules: the valley as a switch}\label{sec:valve}

\subsection{Two modes and the pulse atlas}\label{sec:atlas}

Exact composition makes every schedule of polarisations a certifiable channel. At QR1 we
read out after $n = 88$ rounds, one fewer than the floor, and use two modes. In mode
OPEN the bath is held inside the window at $p_{\mathrm h} = 0.4016$, where the
$88$-round composite is certified PPT (the valve is closed), and pulsed to a lower
polarisation, out of the window, for $k$ consecutive rounds placed at the start (early),
in the middle (centred) or at the end (late) of the schedule; a centred pulse occupies
rounds $a+1$ to $a+k$ with $a = \lfloor (n-k)/2 \rfloor$. In mode CLOSE the bath is held
hot at $p = 0.05$, outside the window, where the $88$-round composite is certified NPT
(the valve is open), and pulsed into the window to break the entanglement.
Table~\ref{tab:atlas} gives the minimal switching pulses.

\begin{table}[t]
\centering
\begin{tabular}{lccc}
\toprule
\multicolumn{4}{l}{mode OPEN: hold $0.4016$, pulse out of the window}\\
pulse polarisation & minimal $k$: early / centred / late & signal at $\kmin{+}4$ (centred) &\\
\midrule
$0.05$  & $32$ / $\mathbf{8}$ / $33$ & $7.9\times10^{-4}$ &\\
$0.113$ & $35$ / $9$ / $35$          & $5.3\times10^{-4}$ &\\
$0.177$ & $38$ / $12$ / $39$         & $4.0\times10^{-4}$ &\\
$0.24$  & $44$ / $18$ / $44$         & $3.1\times10^{-4}$ &\\
$0.303$ & $56$ / $32$ / $57$         & $1.1\times10^{-4}$ &\\
\midrule
\multicolumn{4}{l}{mode CLOSE: hold $0.05$, pulse into the window}\\
pulse polarisation & minimal $k$, centred & & \\
\midrule
$0.4016$ & $35$ & &\\
$0.43$   & $32$ & &\\
$0.556$  & $24$ & &\\
$0.62$   & $21$ & &\\
\bottomrule
\end{tabular}
\caption{Pulse atlas at QR1 ($n = 88$; float64): the minimal pulse length that switches
the composite, from PPT to NPT in mode OPEN and from NPT to PPT in mode CLOSE, by pulse
polarisation and placement, and in mode OPEN the switched signal $-\lmin$ four rounds
above the centred threshold. The bold entry is certified in Sec.~\ref{sec:minimal}.}
\label{tab:atlas}
\end{table}

Three features stand out. Placement matters fourfold: centred pulses switch with $8$
rounds where pulses at either end need $32$ or $33$, an interior optimum in time that
mirrors the interior optimum of the wall in the floor. The hottest pulse, to $p = 0.05$,
is the cheapest, and the cost rises steeply as the pulse polarisation approaches the
window edge at $0.3176$. And the valve is asymmetric: opening costs $8$ rounds with the
hottest pulse, closing $21$ to $35$, between $2.6$ and $4.4$ times more. At QC1 ($n = 80$)
the same atlas is uniformly cheaper, with a minimal centred pulse of $5$ rounds at
$p = 0.05$.

\subsection{The minimal certified pulse}\label{sec:minimal}

The headline entry of the atlas is certified completely. With hold $0.4016$, pulse
polarisation $0.05$ and $n = 88$:
\begin{enumerate}
\item the unpulsed hold and the centred pulses of $k = 1, 2, \dots, 7$ rounds all give a
      composite certified PPT at round $88$;
\item the centred pulse of $k = 8$ rounds gives a composite certified NPT at round $88$,
      with witness quotient $-1.4285084\times10^{-4}$; as a consistency check of the
      ideal property, it is certified NPT at every one of the $88$ rounds;
\item the centred pulse of $k = 7$ rounds is certified NPT at rounds $1$ to $87$ and PPT
      at round $88$, so its crossing lies in the last round interval;
\item the same $8$-round pulse placed at the start gives a composite certified PPT at
      round $88$.
\end{enumerate}
The minimal centred pulse at QR1 is therefore exactly eight rounds, and its dependence on
placement is a certificate rather than a numerical impression. The certificate needs every
shorter pulse, not only the one just below the threshold: in general the end margin need
not be monotone in the pulse length, and a bracket $k_{\mathrm{lo}} < \kmin \le
k_{\mathrm{up}}$ holds only when all pulses up to $k_{\mathrm{lo}}$ are certified not to
switch. Here $k_{\mathrm{lo}} = 7$ and $k_{\mathrm{up}} = 8$.

\subsection{The pulse beats the hold}\label{sec:signal}

The switched negativity is not largest for the longest pulse. Along pulse polarisation
$0.05$ the certified enclosures of $-\lmin$ at round $88$ are
\begin{equation}
\begin{aligned}
  k &= 8:\ 1.4285\times10^{-4}; & k &= 12:\ 7.8712\times10^{-4}; & k &= 30:\ 2.7800\times10^{-3};\\
  k &= 53:\ \mathbf{3.4277926\times10^{-3}}; & k &= 70:\ 3.2338\times10^{-3}; & k &= 88:\ 2.8502660\times10^{-3},
\end{aligned}
\label{eq:landscape}
\end{equation}
with the maximum of the float64 scan at $k = 53$. The endpoint $k = 88$ is the static
protocol at $p = 0.05$,
and it reproduces the certified $\nu_{0.05}$ of QR1 to every printed digit, so the
schedule engine and the static pipeline agree. The comparison is itself certified: the
enclosures at $k = 53$ and $k = 88$ are disjoint, separated by at least
$5.7752661\times10^{-4}$. A centred $53$-round excursion delivers $20$ per cent more
negativity than the static channel at $p = 0.05$, so the best measurement of this valley
is pulsed. Pulses to polarisations closer to the window do worse: at $0.113$, $0.177$ and
$0.24$ the float64 optima lie at $k = 57$, $60$ and $61$, with certified values
$2.9522\times10^{-3}$, $2.2666\times10^{-3}$ and $1.4066\times10^{-3}$.

\subsection{Walking a depth-three valley in time}\label{sec:walk}

D3 has floor $109$ and index exactly $106$ at $p^\ast = 0.5567$. Trading temperature for
time, consider the schedule of $k$ rounds at $p^\ast$ followed by rounds at $p = 0.05$.
Its crossing round $b(k)$ descends the staircase as $k$ grows:
\begin{equation}
  b(k) = 109\ (k < 48),\qquad 108\ (48 \le k < 60),\qquad 107\ (60 \le k < 76),\qquad
  106\ (k \ge 76).
  \label{eq:walk}
\end{equation}
All three steps are certified from both sides and in duration: at each step the
composite is PPT at round $b$, NPT by witness at round $b - 1$, and still NPT at round
$b$ when the pulse is one round shorter. The values between the certified steps are
float64. Each landing is an integer, so this is a death-round readout rather than an
amplitude measurement.

\section{Why the valve opens easily and closes slowly}\label{sec:asym}

Opening the valve at QR1 costs $8$ centred rounds at the best pulse polarisation;
closing it costs $21$ to $35$. This section asks where the factor of $2.6$ to $4.4$
comes from.

\subsection{Two distances}\label{sec:distances}

The two modes start from different distances to the entanglement-breaking boundary. In
mode OPEN the unpulsed composite at round $88$ has
$\lmin = +1.3011396\times10^{-3}$ (certified), a PPT margin that the pulse must remove.
In mode CLOSE the unpulsed composite has $\lmin = -2.8502660\times10^{-3}$ (certified,
equal to $-\nu_{0.05}$ of Table~\ref{tab:targets}), an entanglement margin that the pulse
must remove. Closing starts $2.19$ times further from the boundary. This accounts for
part of the asymmetry and leaves a factor of $1.2$ to $2.0$, depending on the pulse
polarisation, to explain.

\subsection{Per-round rates agree at matched positions}\label{sec:rates}

The natural first hypothesis is kinetic: perhaps rounds of one type move the system
toward the boundary faster than rounds of the other. Along a schedule, let $\lambda(r)$
be $\lmin$ of the composite of the first $r$ rounds, and define the log-margin and its
per-round increment, the \emph{lever},
\begin{equation}
  \ell(r) = \log|\lambda(r)|, \qquad L(r) = \ell(r) - \ell(r-1),
  \label{eq:lever}
\end{equation}
on any stretch of rounds where $\lambda$ keeps its sign. If a pulsed schedule differs from
the hold schedule only inside the pulse block, its \emph{direct lever} at round $r$ is
$\Delta(r) = L_{\mathrm{pulsed}}(r) - L_{\mathrm{hold}}(r)$: the difference between a pulse
round and a hold round at the same position in the circuit.

\begin{table}[t]
\centering
\small
\setlength{\tabcolsep}{4pt}
\begin{tabular}{lcccccc}
\toprule
target, pulse & early hold & mid-pulse & hold, & late hold & pulse / hold, & late / early\\
 & $L(r_1)$ & $L(r_2)$ & $L_{\mathrm{hold}}(r_2)$ & $L(r_3)$ & same round & \\
\midrule
QR1, $k = 12$ & $-0.0224949$ & $-0.0304554$ & $-0.0301817$ & $-0.0780584$ & $1.0091$ & $3.47$\\
 & ($r_1 = 25$) & ($r_2 = 44$) & & ($r_3 = 75$) & & \\
QC1, $k = 9$ & $-0.0238709$ & $-0.0334589$ & $-0.0331837$ & $-0.0993570$ & $1.0083$ & $4.16$\\
 & ($r_1 = 20$) & ($r_2 = 40$) & & ($r_3 = 70$) & & \\
D3, $k = 26$ & $-0.0175757$ & $-0.0253851$ & $-0.0251189$ & $-0.0776393$ & $1.0106$ & $4.42$\\
 & ($r_1 = 25$) & ($r_2 = 55$) & & ($r_3 = 95$) & & \\
\bottomrule
\end{tabular}
\caption{Certified levers along centred OPEN schedules (hold $0.4016$, pulse polarisation
$0.05$; $n = 88$, $80$ and $108$). Each entry is the midpoint of an enclosure of width
below $10^{-13}$. At the same round a pulse round and a hold round differ by about one
per cent, while the lever grows by a factor of $3.5$ to $4.4$ from early to late rounds.}
\label{tab:levers}
\end{table}

Table~\ref{tab:levers} gives certified levers on the OPEN schedules of the three
targets. At the mid-pulse round, the lever of the pulsed schedule and that of the hold
schedule at the same round differ by $0.8$ to $1.1$ per cent; on the CLOSE schedules of
QR1 ($k = 25$, round $44$) the ratio is $0.977$ to $0.990$ across the four window
polarisations (certified). At matched positions the two round types thus move the
log-margin of the running composite at rates that agree to within about two per cent,
while the modes differ in cost by a factor of $2.6$ to $4.4$. What does vary strongly is
position. At every target the lever grows by a factor of $3.5$ to $4.4$ between an early
and a late round of the same schedule, and comparing a pulse round at round $44$ with a
hold round at round $25$ gives a ratio of $1.35$ that is almost entirely this growth.

The direct lever itself is small and depends on the pulse polarisation in opposite ways
in the two modes (Table~\ref{tab:direct}). In mode OPEN, $|\Delta|$ falls from
$2.7\times10^{-4}$ to $1.2\times10^{-4}$ per round as the pulse polarisation approaches
the window edge; across the five QR1 and four QC1 rows it lies between
$1.18\times10^{-4}$ and $2.79\times10^{-4}$. In mode CLOSE, $\Delta$ grows from
$2.9\times10^{-4}$ to $7.0\times10^{-4}$ as the pulse moves deeper into the window. Two
features show that the cost is not set by this local effect. The mode with the larger
direct lever is the expensive one. And in both modes the direct lever has the sign
opposite to the pulse's eventual effect: relative to the hold, an OPEN pulse round brings
the running composite slightly faster toward the boundary ($\Delta < 0$) and a CLOSE
pulse round slightly more slowly ($\Delta > 0$), yet the first pulse keeps the end
composite entangled and the second breaks it. A pulse acts through how it changes the evolution under the hold rounds that
follow, not through a local push.

\begin{table}[t]
\centering
\small
\begin{tabular}{lcc@{\qquad}lc}
\toprule
\multicolumn{3}{l}{mode OPEN ($k = 12$; QR1 at $r = 44$, QC1 at $r = 40$)} & \multicolumn{2}{l}{mode CLOSE ($k = 25$; QR1 at $r = 44$)}\\
pulse $p$ & QR1 & QC1 & pulse $p$ & QR1\\
\midrule
$0.05$  & $-2.7377\times10^{-4}$ & $-2.7906\times10^{-4}$ & $0.4016$ & $+2.9273\times10^{-4}$\\
$0.113$ & $-2.5491\times10^{-4}$ & $-2.5953\times10^{-4}$ & $0.43$   & $+3.3632\times10^{-4}$\\
$0.177$ & $-2.2224\times10^{-4}$ & $-2.2608\times10^{-4}$ & $0.556$  & $+5.6556\times10^{-4}$\\
$0.24$  & $-1.7679\times10^{-4}$ & $-1.7974\times10^{-4}$ & $0.62$   & $+7.0436\times10^{-4}$\\
$0.303$ & $-1.1816\times10^{-4}$ & & & \\
\bottomrule
\end{tabular}
\caption{Certified direct levers $\Delta$ at a mid-pulse round of centred schedules: hot
pulses against the $0.4016$ hold (OPEN) and window pulses against the $0.05$ hold
(CLOSE).}
\label{tab:direct}
\end{table}

\subsection{How the end margin responds}\label{sec:response}

\begin{figure}[t]
\centering
\includegraphics[width=0.94\textwidth]{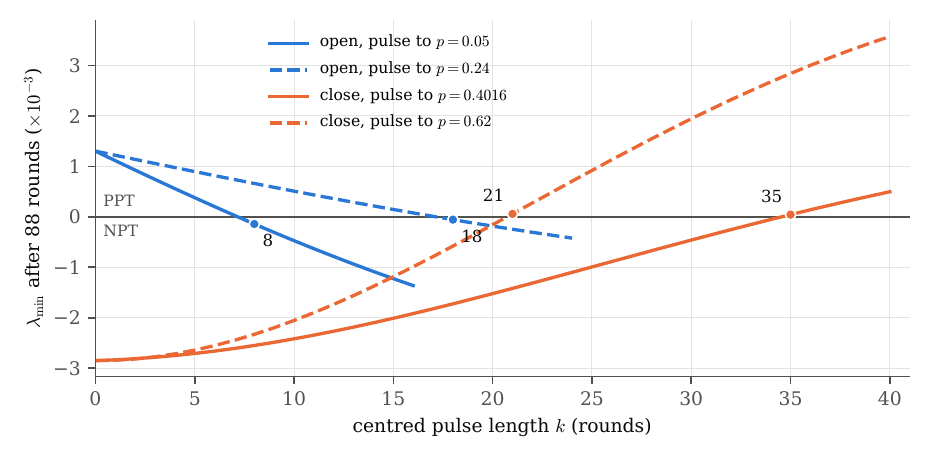}
\caption{End margin of the $88$-round composite at QR1 against the length of a centred
pulse (float64). Mode OPEN (blue): hold $0.4016$, pulses to $p = 0.05$ (solid) and
$0.24$ (dashed); mode CLOSE (orange): hold $0.05$, pulses to $0.4016$ (solid) and $0.62$
(dashed). Above zero the composite is PPT (entanglement broken), below zero NPT
(entangled). Dots mark the minimal switching pulse; the values at $k = 0$ and the switch at
$k = 8$ (solid blue) are certified.}
\label{fig:response}
\end{figure}

The quantity that decides the switch is $\lmin$ of the full composite at the read-out
round, the end margin, as a function of the pulse length (Fig.~\ref{fig:response}). In
mode OPEN at pulse polarisation $0.05$ the response is almost linear: each centred hot
round moves the end margin by $-1.89\times10^{-4}$ for the first round and by
$-1.80\times10^{-4}$ on average over the first eight, and $8$ rounds cover the
$1.30\times10^{-3}$ that must be removed. In mode CLOSE the response is collective. A
single window round at polarisation $0.4016$ inserted in the hot hold moves the end
margin by only $+1.5\times10^{-5}$, about twelve times less than a single hot round in
mode OPEN; the increment per added round rises to $+6.9\times10^{-5}$ at the tenth round
and about $+1.0\times10^{-4}$ from the twentieth on, and $35$ rounds are needed to cover
$2.85\times10^{-3}$. Deeper window pulses build up faster, and at polarisation $0.62$ the
cost falls to $21$ rounds (all float64).

The asymmetry thus factorises into the distance ratio of Sec.~\ref{sec:distances},
$2.19$, and the ratio of the average response needed per pulse round, the margin to be removed divided by
$\kmin$,
$(1.30\times10^{-3}/8)/(2.85\times10^{-3}/35) = 2.0$ at window polarisation $0.4016$ and
$1.2$ at $0.62$: together they give the observed cost ratios $35/8$ and $21/8$. The
factorisation is an identity; its content is the shape of the two responses. Hot rounds
added to a window hold act almost independently of one another, while window rounds
added to a hot hold change the end margin little until enough of them act together.

Placement shapes the response as strongly as the mode does. A single hot round at the
start or the end of the schedule moves the end margin by $+2.3\times10^{-5}$ and
$+2.6\times10^{-5}$, in the wrong direction, where the same round in the middle moves it
by $-1.9\times10^{-4}$; blocks at either end switch only at $32$ or $33$ rounds
(float64). Where a pulse acts in the circuit decides not only how much it does but in
which direction.

\subsection{The crossing and the lever}\label{sec:graze}

\begin{figure}[t]
\centering
\includegraphics[width=0.98\textwidth]{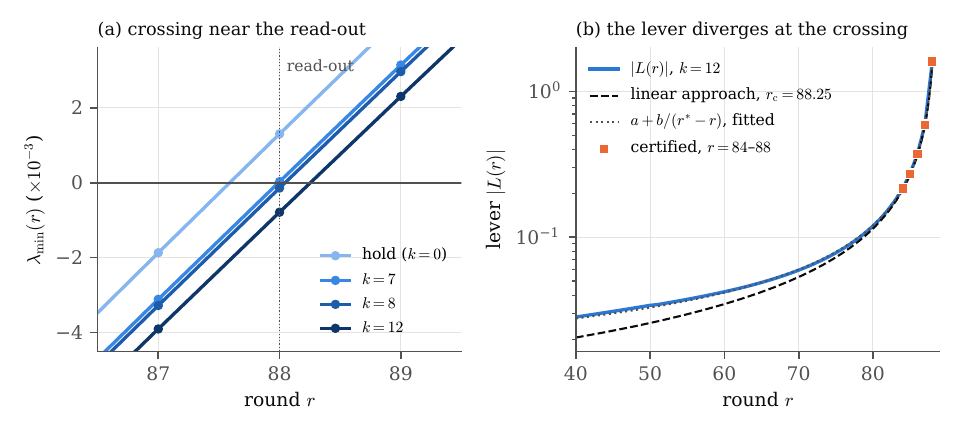}
\caption{(a) $\lmin$ of the composite of the first $r$ rounds at QR1 near the read-out,
for the unpulsed hold and centred pulses of $7$, $8$ and $12$ rounds to $p = 0.05$
(float64; points at integer rounds; all four schedules are continued with hold rounds
beyond round $88$). One round of pulse moves the crossing past round
$88$. (b) The lever $|L(r)|$ on the $k = 12$ schedule (float64), the parameter-free law of
Eq.~\eqref{eq:linlaw} for a linear approach crossing at $88.2548$ (dashed), the effective
law of Eq.~\eqref{eq:graze} (dotted), and the certified levers at rounds $84$ to $88$
(squares).}
\label{fig:graze}
\end{figure}

Near the read-out the natural coordinate is the crossing round rather than the margin.
Interpolating $\lambda(r)$ linearly between the last NPT round and the first PPT round
(float64), the crossing of the unpulsed hold at QR1 lies at round $87.59$; centred hot
pulses delay it to $87.99$ at $k = 7$, $88.05$ at $k = 8$ and $88.25$ at $k = 12$
(Fig.~\ref{fig:graze}a). Near threshold each pulse round delays the crossing by $0.05$ to
$0.06$ rounds, and the switch happens when the crossing passes the read-out round. The
certificates of Sec.~\ref{sec:minimal} make this exact at the threshold: the $k = 7$
crossing lies in the last round interval, and the $k = 8$ composite is still NPT at
round $88$.

Because $\lambda$ approaches zero almost linearly in $r$, the lever near the crossing
follows from the crossing round alone. If $|\lambda(r)| = s\,(r_{\mathrm c} - r)$ for some
slope $s$, then
\begin{equation}
  |L(r)| \;=\; \ln\!\Bigl(1 + \frac{1}{r_{\mathrm c} - r}\Bigr)
  \label{eq:linlaw}
\end{equation}
exactly, whatever $s$, and the lever diverges as the crossing is approached. With
$r_{\mathrm c} = 88.2548$, the interpolated crossing of the $k = 12$ schedule, and no fitted
parameter, Eq.~\eqref{eq:linlaw} reproduces the certified levers at rounds $84$ to $88$,
$0.2155$, $0.2723$, $0.3715$, $0.5906$ and $1.6010$, to within $2.1$ per cent
(Fig.~\ref{fig:graze}b). Further from the crossing the approach is not linear, and the
lever over the whole schedule is better described by an effective law,
\begin{equation}
  |L(r)| \;\approx\; a + \frac{b}{r^\ast - r},
  \label{eq:graze}
\end{equation}
with $a = 0.0078$, $b = 0.972$ and $r^\ast = 88.62$ from a fit in logarithmic space with
coefficient of determination $0.9995$ (float64). The same fit applied to an exactly linear
approach that crosses at $88.2548$ returns $b = 0.987$ and $r^\ast = 88.63$: the offset of
$r^\ast$ from the crossing is an effect of discreteness, and so is about half of the
departure of $b$ from $1$. The fitted $r^\ast$ therefore does not locate the crossing; the
interpolated crossing does.

The divergence explains why late rounds dominate the log-margin: the lever grows
severalfold toward the read-out (Table~\ref{tab:levers}) and without bound at the
crossing. It does not by itself explain why a block at the end of the schedule acts in
the wrong direction, why the direct lever has the opposite sign to the eventual effect,
or why the response in mode CLOSE builds up collectively; those are properties of how a
block of rounds changes the composite affine pair, and we leave them open.

\subsection{A phenomenological budget law}\label{sec:budget}

A rule of thumb ties the minimal pulse to two local quantities: the end margin $m$ to be
removed and the mean direct lever $\bar\Delta$ over the pulse block on the minimal
schedule,
\begin{equation}
  \kmin \;\approx\; \frac{2m}{|\bar\Delta|}.
  \label{eq:budget}
\end{equation}
Over the nine OPEN rows of QR1 and QC1 the ratio of the true $\kmin$ to this estimate
lies between $0.83$ and $1.29$; over the four CLOSE rows it lies between $1.29$ and
$2.49$; and for pulses to $p = 0.05$ at the four targets with floors $60$ to $109$ it
lies between $0.73$ and $1.11$ (float64). At the floor-$28$ target of the search, where
$\kmin = 1$, the estimate is $0.25$, below the resolution of an integer observable. The
law is phenomenology, not a mechanism, and Sec.~\ref{sec:response} shows why it fails on
the CLOSE side: it uses one local rate where the response there is collective.

\begin{remark}[D3]\label{rem:d3}
The hold trajectory of D3 crosses one round before its read-out: at hold $0.4016$ the
composite is certified NPT at round $106$ ($\lmin = -1.405\times10^{-3}$) and PPT at
rounds $107$ and $108$ ($+1.240\times10^{-3}$ and $+3.861\times10^{-3}$). At QR1 and QC1
the hold crosses in the last round interval. Comparisons across targets that treat the
read-out as the crossing round should carry this offset.
\end{remark}

\section{The coherent tilt}\label{sec:tilt}

The analysis so far used diagonal thermal ancillas. Its extension to an arbitrary ancilla
Bloch vector $\mathbf r$ is exact, because the affine pair remains polynomial in the
components of $\mathbf r$, and the general builder reproduces the diagonal one
identically at $\mathbf r = (0, 0, p)$. Fix the population at the hold value,
$r_z = 0.4016$, so that the populations of the ancillas, and with them their energy, do
not change, and tilt the Bloch vector (Table~\ref{tab:tilt}).

\begin{table}[t]
\centering
\footnotesize
\setlength{\tabcolsep}{4pt}
\begin{tabular}{lll}
\toprule
drive (all at $r_z = 0.4016$, $n = 88$) & result & note\\
\midrule
$r_y$ tilt, every round & valve opens at $r_y^\ast \in (0.0119,\ 0.0121)$, certified & a $1.2$ per cent tilt\\
$r_x$ tilt, every round & threshold $0.329$ (float64) & $27$ times stiffer\\
$r_y = 0.05$, every round & signal $3.8601657\times10^{-3}$, certified & beats the best thermal pulse\\
$r_y = 0.1$, every round & signal $7.0340090\times10^{-3}$, certified & $2.5$ times the static signal\\
$r_y = 0.2$, centred $k = 26$ & signal $8.5895444\times10^{-3}$, certified & $3$ times the static signal\\
$r_y = 0.2$, centred $k = 2$ & switches (NPT); $k = 1$ does not (PPT), certified & a two-round switch\\
\bottomrule
\end{tabular}
\caption{Coherent tilt at QR1 with the population fixed at $r_z = 0.4016$. Signals are
$-\lmin$ of the $88$-round composite. A centred $k$ means that the tilted ancilla replaces
the hold ancilla for $k$ rounds in the middle of the schedule. Certified entries are
two-sided enclosures or sign certificates.}
\label{tab:tilt}
\end{table}

The comparison with the thermal route is certified: the signal under a sustained tilt
$r_y = 0.1$ exceeds the best thermal pulse ($k = 53$, Sec.~\ref{sec:signal}) by at least
$3.6062\times10^{-3}$. A tilt seven times smaller than the population change from the
hold value to the window edge opens the valve; a sustained tilt $r_y = 0.1$ raises the
static signal $2.5$-fold and a centred $26$-round tilt at $r_y = 0.2$ triples it, all
without changing the populations. The anisotropy is informative: $Y$, the
soft axis of the tilt, is also the bath direction along which the static index of the
loop's reference circuit is largest, $14$ against $3$ for a $Z$-polarised bath at zero
temperature \cite{PaperII}, so the sensitivity to ancilla coherence and the sensitivity
to bath direction point the same way. A coherent ancilla is not a classical mixture of
basis states, so a tilt cannot be programmed by sampling basis states as a thermal
ancilla can; it needs one rotation per ancilla per round, and its exactness becomes a
calibration question.

\section{Noise fragility}\label{sec:noise}

Within the exact per-round depolarising model of Sec.~\ref{sec:affine} the following are
certified at QR1: the switched signal of the centred $k = 12$ pulse is dead at
$\eta = 10^{-4}$; the static signal $\nu_{0.05}$ is alive at $\eta = 10^{-4}$ and dead at
$1.4\times10^{-4}$ and at $2\times10^{-4}$; and the signal of the centred $53$-round pulse
is alive at $\eta = 1.4\times10^{-4}$.

The thresholds follow from the noiseless signal alone. If the composite affine pair is
contracted uniformly, $(A, \mathbf c) \to s\,(A, \mathbf c)$, then
$H \to s\,H + \tfrac{1-s}{4}\,\mathbb{I}$, every eigenvalue moves linearly toward
$\tfrac14$, $\lmin = \tfrac14 - s\,(\tfrac14 + \nu)$ with $\nu$ the signal at $s = 1$, and
the signal dies at $s = 1/(1 + 4\nu)$. Per-round noise contracts $A$ by exactly
$(1-\eta)^n$ but $\mathbf c$ only approximately so, the contribution of round $j$ being
contracted by $(1-\eta)^{n-j+1}$. Even so, for the static protocol and for the seven
centred pulses tested, of $8$ to $80$ rounds, the float64 thresholds satisfy
\begin{equation}
  (1 - \eta^\ast)^{n} \;=\; \frac{1}{1 + 4\nu}, \qquad\text{so that}\qquad
  \eta^\ast \;\approx\; \frac{4\nu}{n},
  \label{eq:noise}
\end{equation}
to within $0.04$ per cent. The static threshold is $\eta^\ast = 1.288\times10^{-4}$. The
$k = 12$ pulse dies at a noise level $3.6$ times lower, $3.57\times10^{-5}$, because its
signal is $3.6$ times smaller, and the $53$-round pulse survives to a level $20$ per cent
higher, $1.55\times10^{-4}$, because its signal is $20$ per cent larger. The eight-round
switch, with a signal of only $1.4\times10^{-4}$, dies already at $6.5\times10^{-6}$.
At QR1 the fragility is therefore that of the signal itself, about $88$ rounds acting on
a margin of order $10^{-3}$ or less, and pulsing neither causes nor cures it. We have
tested Eq.~\eqref{eq:noise} only at QR1. A measurement of the static signal or of the
strongest pulses must reach an effective per-round error of order $10^{-4}$, and one of
the minimal switch an error some twenty times smaller, by hardware or by extrapolation.
What the pulsed schedules add is not robustness but discriminating dials:
the position, polarisation and length of the pulse change the certified effect, while
noise that does not depend on them cannot.

\section{What is certified, and what is not}\label{sec:claims}

\paragraph{Certified.} Two-sided enclosures or proven signs and integers in ball
arithmetic, re-derived by the scripts of Appendix~\ref{app:files}: the floors, window
sign patterns and contrasts of the targets of Table~\ref{tab:targets} and of the three
search anchors, and the floor and in-window index of D3; the true window edges of QR1;
the minimal centred pulse of exactly eight rounds, with every shorter centred pulse and
the early eight-round pulse failing; the signal enclosures of Eq.~\eqref{eq:landscape},
the separation of the $53$-round pulse from the static protocol, and the signal values at
the optima for pulse polarisations $0.113$, $0.177$ and $0.24$; the three steps of the
D3 walk; the end margins of Sec.~\ref{sec:distances} and Remark~\ref{rem:d3}; the levers
and direct levers of Tables~\ref{tab:levers} and~\ref{tab:direct} and the final-round
levers of Sec.~\ref{sec:graze}; the rejection of the planted false claims; the tilt
threshold, the three tilt signals, the two-round switch and the tilt-beats-pulse
separation; and the noise statements of Sec.~\ref{sec:noise}.

\paragraph{Float64, labelled.} The search statistics and Table~\ref{tab:wall}; the
valley abundances and depth fractions; the window of D3; the pulse atlas
(Table~\ref{tab:atlas}) apart from its certified entry; the response curves and crossing
rounds of Secs.~\ref{sec:response} and~\ref{sec:graze}; the location in $k$ of the
signal optima; the effective law Eq.~\eqref{eq:graze} and the comparison of
Eq.~\eqref{eq:linlaw} with the certified levers; the budget-law ratios; the values between the certified steps of
the D3 walk; the $r_x$ threshold; and the noise thresholds and their agreement with
Eq.~\eqref{eq:noise}.

\paragraph{Empirical, with certified extreme points.} The absence of valleys with
$C \ge 0.05$ in $33$ million draws, the interior maximum of the wall, and the conjecture
of Sec.~\ref{sec:wall} rest on the search, not on a proof.

\paragraph{Not claimed.} We do not claim that any of these signatures is measurable on
present hardware (Sec.~\ref{sec:conclusion} gives the cost), that the effective law is
exact or locates the crossing, that the budget law is a mechanism, anything about the
entropy flow of the loop, or that the hold value $0.4016$ is special.

\section{Discussion and outlook}\label{sec:conclusion}

A thermal valley turns a static curiosity into a switch: held inside its window, a
collision channel breaks all entanglement by its read-out round, and a short, centred
excursion out of the window restores it. At the reference target the threshold is
exactly eight rounds, the best pulse beats the static protocol by twenty per cent, a
depth-three valley can be walked down its staircase by pulse duration, and a small
coherent tilt opens the switch at fixed population. The switch opens more easily than it
closes, and the reason is not a difference of local rates, which agree at matched
positions to within about two per cent. Closing starts more than twice as far from the
boundary, and the response to window rounds builds up only when enough of them act
together, while hot rounds act almost one by one. Near the read-out the lever of the
log-margin is that of a linear approach to the boundary, and the switch is the moment at
which a pulse moves the crossing past the read-out round.

The cost of seeing any of this is high. Version~3 of Ref.~\cite{PaperV} estimates that a
margin of $3\times10^{-3}$ costs about $4\times10^{5}$ shots per setting for each of the
$18$ settings of its tomography (version~2 had estimated $3\times10^{6}$ from an assumed
error model), so a sixteen-point sweep of a valley needs of order $10^{8}$ shots, on
circuits of $80$ to $110$ rounds with mid-circuit reset. Together with the noise
requirement of Sec.~\ref{sec:noise}, this places valley experiments beyond current
devices. Engineered thermal baths have been swept between about $0.08$ and $0.36$~K
\cite{Ronzani2018}. For a $6$~GHz mode the hold polarisation corresponds to $338$~mK,
inside that range, and the true window of QR1 to $93$ to $438$~mK, whose hot end lies
beyond it, as do the pulse temperatures of mode OPEN, from $0.46$~K at $p = 0.303$ to
$2.9$~K at $p = 0.05$. On a processor, where a thermal ancilla can be programmed by
sampling basis states, these temperatures are settings rather than bath conditions.

Several questions are sharper for this analysis. The boundary-geometry conjecture of
Sec.~\ref{sec:wall} asks for a bound on contrast in terms of distance to the
entanglement-breaking set. The response of the end margin to a block of rounds, linear in
one mode and collective in the other, and of opposite sign at the ends of the schedule,
should follow from how a block changes the composite affine pair, and a theory of it
would replace the phenomenological budget law. The optimal schedule is a small control
problem whose value function is certifiable pointwise, and the affine pair of a mixed
bath $\sum_i w_i\Phi_{p_i}$ is affine in the weights, so its index is certifiable with
the same machinery.
Finally, in dimension $d \ge 3$ the partial transpose no longer decides separability, and
a valve there would need certified separability rather than certified negativity.

\section*{Acknowledgements}
\addcontentsline{toc}{section}{Acknowledgements}

The computations used the open-source libraries Arb (through python-flint), NumPy and
Matplotlib.

\paragraph{Funding} No external funding was received for this work.

\paragraph{Use of AI tools}
This work was prepared with Claude (Anthropic), used through the Claude desktop
application in August to October 2026 (this version with Claude Opus 5.5). It was used to
write and run the search, certification and analysis scripts, to draft and revise the
manuscript text, and to check bibliographic details. The author set the research
questions, chose the claims and reviewed all text and code. The deposited scripts reproduce
every computed result in the paper from the stated inputs and, for the search
statistics, from the deposited search output.

\paragraph{Data and code availability} The ancillary files of the arXiv record of this
paper contain the certified channel model, the search chassis and certification
pipeline, the schedule engine and general-ancilla builder, the complete valley lists and
draw tallies of the three searches, the certificate files, the float64 atlases, the
scripts that re-derive every certified number, and the script that draws the figures
(Appendix~\ref{app:files}). The certification library from which the primitives derive
is archived at Zenodo \cite{Software}.

\appendix

\section{Certification methodology}\label{app:cert}

All certificates use Arb ball arithmetic through python-flint at $256$ bits, with
parameters entered as described in Sec.~\ref{sec:cert}. The round unitary is built from closed-form sines and
cosines of the four angles, using $(Z\otimes Y)^2 = \mathrm{SWAP}^2 = \mathbb{I}$; the
affine pair follows by tracing out the ancillas, and composites by
Eq.~\eqref{eq:compose}, every step carrying a rigorous radius. Signs of $\lmin$ of the
$4\times4$ partially transposed Choi matrix $H$ are established by one-sided primitives:
negativity by an explicit witness vector $v$ with $(v^\dagger Hv)/(v^\dagger v)$ certified
negative, positivity by Sylvester's criterion on the leading principal minors, each
certified strictly positive. Two-sided enclosures combine the Rayleigh upper bound at a
floating-point eigenvector with Sylvester's criterion on $H - s\,\mathbb{I}$, $s$ being
backed off geometrically from the floating-point value until the shifted certificate
passes; a pass at any $s$ is rigorous. An enclosure that straddles zero is a failed sign
certificate. Levers are differences of logarithms of such enclosures, computed in ball
arithmetic on stretches where the sign of $\lambda$ is certified constant.

Searches and scans are float64 (NumPy). The closed-form floor used as a prescreen is
checked against the Choi-matrix floor in every run, and every kept circuit's
sixteen-point profile is checked for sign consistency between the index profile and the
margins wherever $|\lmin| > 10^{-9}$; a mismatch aborts the shard. Floating-point
routines propose witnesses and explore; nothing quoted as certified rests on them.

\section{Ancillary files and re-derivation}\label{app:files}

{\sloppy The ancillary directory of the arXiv record contains the files of
Table~\ref{tab:files}; the files of the earlier version are carried over unchanged. The
script \texttt{paperV\_checks.py} re-derives from the primitives alone the floors and
window sign patterns of the targets and anchors, two-sided enclosures of the binding
contrast term of QR1 and of the floor-$3$ reference valley, the window edges, the signal
enclosures, the D3 walk, the tilt and noise certificates and the float64 statistics of
the searches, reading the certificate files only to compare; for the other targets and
anchors it recomputes the contrast from Rayleigh quotients and matches the certified
values. It also covers the measurement-design results of the earlier version
\cite{PaperV}, whose certificate files are kept so that it runs unchanged. The two-sided
contrast certificates are re-issued by the certification pipeline: run on the deposited
valley list of a search (with \texttt{-{}-out} set, so that the deposited certificate file
is not overwritten), \texttt{certify\_scan4.py} certifies its three strongest true
returns and its strongest cliff, which are QR1, TR2, TR3 and QCL for
\texttt{valleys\_probe120.csv} and include the two anchors for the other two lists. The
minimal-pulse certificates for $k = 7$, $8$ and the early $8$-round pulse are reproduced
by \texttt{pulse\_engine.py} when run as a script, and for every $k \le 8$ by
\texttt{valve\_asymmetry\_checks.py}, which also certifies the contrast of QC1 (not a
search winner) two-sidedly, the margins at the window-edge brackets, the end margins, the
levers and direct levers of Sec.~\ref{sec:asym} and the noise comparison of
Sec.~\ref{sec:noise}, rejects the planted false claims, recomputes the float64
quantities of Secs.~\ref{sec:asym} and~\ref{sec:noise}, including the budget-law ratios,
the comparison with Eq.~\eqref{eq:linlaw} and Eq.~\eqref{eq:noise}, and recomputes the
deposited float64 tables (the pulse atlas, the signal optima, the D3 walk and the tilt
thresholds), the power-law extrapolation of Sec.~\ref{sec:wall} and the shot arithmetic
of Sec.~\ref{sec:conclusion}. The
check scripts need NumPy and python-flint ($0.9.0$ tested; Matplotlib for the figures),
run in under half a minute, and exit with a non-zero status if any check fails; the
search chassis \texttt{scan4.py} reproduces the searches in hours.\par}

\begin{table}[!ht]
\centering
\footnotesize
\begin{tabular}{>{\raggedright\arraybackslash}p{0.42\textwidth}>{\raggedright\arraybackslash}p{0.52\textwidth}}
\toprule
file & contents\\
\midrule
\texttt{certified\_model.py} & certified channel model: exact affine pair, Choi primitives\\
\texttt{scan4.py}, \texttt{certify\_scan4.py} & gated search chassis; certification pipeline\\
\texttt{pulse\_engine.py} & schedule engine and general-ancilla builder; reproduces the minimal-pulse, signal and tilt certificates when run\\
\texttt{paperV\_checks.py} & re-derivation of the certificates of Secs.~\ref{sec:valley}, \ref{sec:valve}, \ref{sec:tilt} and~\ref{sec:noise} (see text)\\
\texttt{valve\_asymmetry\_checks.py} & contrast of QC1; window-edge margins; minimal pulse for every $k \le 8$; certificates and float64 checks of Secs.~\ref{sec:asym} and~\ref{sec:noise}; recomputation of the deposited float64 tables\\
\texttt{make\_figures.py} & Figs.~\ref{fig:response} and~\ref{fig:graze}\\
\texttt{scan4\_certificates.json}, \texttt{probe60\_certificates.json}, \texttt{probe120\_certificates.json} & certified winners of the three searches, including the targets of Table~\ref{tab:targets}\\
\texttt{valleys\_scan4.csv}, \texttt{valleys\_probe60.csv}, \texttt{valleys\_probe120.csv} & complete float64 valley lists ($10{,}628$ rows)\\
\texttt{tally\_scan4.tsv}, \texttt{tally\_probe60.tsv}, \texttt{tally\_probe120.tsv} & draw and floor tallies ($33$ million draws)\\
\texttt{pulse\_certificates.json}, \texttt{signal\_plane\_certificates.json}, \texttt{signal\_plane\_f64.json} & minimal pulse; signal enclosures of Eq.~\eqref{eq:landscape}; float64 signal plane\\
\texttt{register2\_certificates.json}, \texttt{register2\_f64.json} & window-edge brackets, D3 walk, tilt threshold, noise flags; float64 companions\\
\texttt{tilt\_pulse\_certificates.json}, \texttt{tilt\_pulse\_f64.json} & Table~\ref{tab:tilt}\\
\texttt{valley\_noise\_certificates.json} & noise statements of Sec.~\ref{sec:noise} at $\eta = 10^{-4}$ and $2\times10^{-4}$\\
\texttt{pulse\_atlas\_qr1\_f64.json} & Table~\ref{tab:atlas}, mode OPEN\\
\texttt{valve\_asymmetry\_results.json} & output of \texttt{valve\_asymmetry\_checks.py}, including the noise statements at $\eta = 1.4\times10^{-4}$\\
\texttt{edge\_noise\_fix.json} & bisection of the window edges\\
\texttt{objA\_margins\_certificates.json}, \texttt{noise\_certificates.json}, \texttt{compile\_verify.json} & measurement-design certificates of the earlier version, read by \texttt{paperV\_checks.py} (not used in the text)\\
\texttt{ptable\_crossover\_30dig.json} & auxiliary table of $p(T)$ from the earlier version (not used)\\
\bottomrule
\end{tabular}
\caption{Ancillary files.}
\label{tab:files}
\end{table}

\section{Relation to earlier versions}\label{app:versions}

Versions~1 and~2 of arXiv:2609.09350 \cite{PaperV} contained the valley search, the pulse
results and the tilt in an earlier form; version~3 of that record is a measurement
protocol and keeps only a short summary of them. Relative to version~2 this paper omits
the measurement design, compilation and staircase targets, which belong to version~3;
certifies the minimal centred pulse against every shorter pulse, where version~2
certified the unpulsed hold, the centred pulses of $7$ and $8$ rounds and the early
$8$-round pulse; gives the cost of mode CLOSE by pulse polarisation; corrects the
statements that sustained tilts triple the static signal, that D3 is the first
certified depth-three valley, and that pulsing costs a factor of about $3.6$ in noise
threshold, which holds only for the $12$-round pulse; and adds Sec.~\ref{sec:asym} and
Eq.~\eqref{eq:noise}.

\FloatBarrier

\end{document}